# Physical properties of spinel-type superconductors $CuRh_2S_4$ and $CuRh_2Se_4$: A DFT study

Md. Ibrahim Kholil and Md. Tofajjol Hossen Bhuiyan*

Department of Physics, Pabna University of Science and Technology, Pabna-6600, Bangladesh

**A B S T R A C T**

The structural, elastic, electronic, Vickers-Hardness, vibrational, Optical and thermodynamical properties of potentially technologically significant superconductors $CuRh_2S_4$ and $CuRh_2Se_4$ have calculated using density functional theory (DFT) with CASTEP code. The calculated lattice parameters and other properties have compared with available experimental values and found good agreement with them. The mechanical stability found for $CuRh_2S_4$ and $CuRh_2Se_4$ under Born stability conditions. Pugh's ratio indicates the both are ductile and Poisson's ratio reveals the brittle nature. The valence band and conduction bands overlapped each other at the Fermi level indicates the metallic conductivity of $CuRh_2S_4$ and $CuRh_2Se_4$. The density of states shows that S-3p and Se-4p states are more effective at the Fermi level. The charge density difference maps indicates the Cu-Rh bonds are stronger than Cu-S. The overall higher discussion notifies that the chemical bonding can be designated as an effective anisotropic connection between ionic, covalent and metallic interactions for $CuRh_2S_4$ and $CuRh_2Se_4$. The Vickers-Hardness indicates the soft material with comparing to Diamond and suitable to use wires and ribbon cables. The electron- and hole-like sheets make the complex multisheet Fermi surface of $CuRh_2S_4$. The different optical functions are also observed clearly. The absorption spectra indicate that $CuRh_2S_4$ is more suitable to use in solar cell rather than $CuRh_2Se_4$. The reflectivity spectrum shows that these compounds have the potential to be used a reflector material. Debye temperature indicates that $CuRh_2S_4$ and $CuRh_2Se_4$ should have advantages to use as a thermal barrier coating (TBC) material. The electron-phonon coupling constant indicates the phonon-mediated medium coupled BCS superconductors. The obtained potential results in present calculation could provide a significant movement for future studies.

---

* Corresponding author.
E-mail addresses: Ibrahim.physics20@gmail.com (Md. I. Kholil); thbapon@gmail.com (Md. T. H. Bhuiyan)

# 1. Introduction

Chalcogenide spinel compounds show wide variety of attractive physical properties such as the magnetic ordering [1], the metal-insulator transition [2,3], and superconductivity [4]. Further, spinel compounds have potential of the technological applications. Due to this attractive physical properties spinel compunds have gained much interest and great attention of researchers. The general formula of chalcogenide spinels is $AB_2X_4$, where A and B are the transition metals and X is the chalcogen [5]. The sulpo- and selenospinels $CuRh_2S_4$ and $CuRh_2Se_4$ exhibit an extensive variety of electrical and magnetic properties such as ferromagnetic and antiferromagnetic [6]. These spinels have normal cubic structure where the Cu atoms occupy the A (tetrahedral) sites and the Rh atoms occupy the B (octahedral) sites [4]. Further, they exhibit metallic conduction and temperature-independent susceptibility. The superconductivity of the ternary sulfo- and selenospinels $CuRh_2S_4$ and $CuRh_2Se_4$ have been found at transition temperature 4.70 K and 3.48 K, respectively [4]. T. Hagino et al. predicted the lattice constant $a = 9.787$ Å and 10.269 Å for $CuRh_2S_4$ and $CuRh_2Se_4$, respectively [4].

Lotgering and Van stapele discuss about electrical and magnetic properties of $CuRh_2S_4$ and $CuRh_2Se_4$ and reported that these compounds show superconducting transition between 3 and 4 K [7-9]. M. Ito and his co-workers investigated the magnetic properties of chalcogenide spinel superconductor $CuRh_2S_4$ under pressure and evaluated the pressure dependence of the superconducting parameters [5]. Shelton et al. investigated the pressure dependence transition temperature ($T_c$) up to $P$=2.2 GPa for the spinels $LiTi_2O_4$, $CuRh_2Se_4$, and $CuRh_2S_4$ [10]. Due to enhancement of the Debye temperature ($\theta$) they predicted that the superconducting transition temperature ($T_c$) increase with pressure ($P$). M. Ito et al. calculated electric resistivity under pressure and predicted the phenomenon of the pressure-induced transition of $CuRh_2S_4$ from a superconductor to an insulator [11]. The highest superconductivity was found in the spinel isostructural compound $LiTi_2O_4$ [12]. T. Oda et al. investigated the band structure, density of states, Fermi surface and charge density of sulphide spinels $CuM_2S_4$ (M =Co, Rh, Ir) [13]. G.L Hart et al. also investigated the band structure of $Cu_{1-x}Ni_xRh_2S_4$ and $CuRh_2Se_4$ [14].

In this paper we have tried to investigate the details physical properties of $CuRh_2S_4$ and $CuRh_2Se_4$ theoretically. Though some theoretical work found in ref. [13] and [14] but still absent about the elastic, vibrational, Vickers-Hardness, optical and thermodynamic properties. Furthermore, considerable progresses have been made after details physical properties calculation of these two spinel superconductors due to rich physical properties. That's why we have investigated the structural, elastic, electronic, bonding, vibrational, optical and thermodynamic properties and electron-phonon coupling constant by using the first principles method base on the density functional theory with CASTEP code.

Finally, the remaining parts of this article are arranged as follows: A brief description of computational method shown in section 2, Investigated results and its related discussion are shown in section 3 and the summary of this work are displaced in section 4.

## 2. Computational methods

The present first principle calculations have been performed using the density functional theory (DFT) based on Cambridge Serial Total Energy Package (CASTEP) computer program [15-18]. The electronic exchange-correlation interaction has been treated by utilizing the generalized gradient approximation (GGA) within the scheme described by Perdew-Burke-Ernzerhof (PBE) [18]. The interactions between ions and electrons are represented with Vanderbilt-type ultrasoft pseudopotentials for Cu, Rh, S and Se atoms [19]. The valence electron configurations of $CuRh_2S_4$ and $CuRh_2Se_4$ superconductors have considered Cu-$3d^{10} 4s^1$, Rh-$4d^8 5s^1$, S-$3s^2 3p^4$ and Se-$4s^2 4p^4$, respectively for pseudo atomic calculations. For all the physical properties calculations have used a plane-wave cutoff energy of 350 eV. Monkhorst-Pack scheme have used to generate 8×8×8 $k$-point grids for the sampling of the Brillouin zone [20]. The Broyden-Fletcher-Goldfarb-Shanno (BFGS) minimizations have used to perform the structural optimizations [21]. In the case of geometry optimization the convergence tolerance have selected as follows: the total energy convergence value is within $2.0×10^{-5}$ eV/atom, the maximum Hellmann-Feynaman force is within 0.05 eV/Å; the maximum displacement is within 0.002 Å; the maximum stress is within 0.01 GPa; and the maximum iterations is within 100. The elastic constants of $CuRh_2S_4$ and $CuRh_2Se_4$ have investigated by the stress-strain method [22]. The maximum strain amplitude have elected to 0.003. The criteria for convergence tolerance to evaluate the elastic constants have used to $4.0× 10^{-6}$ eV/atom for the total energy, 0.01 eV/Å for maximum force and $4.0× 10^{-4}$ Å for maximum displacement. The four numbers of steps have selected for each strain.

## 3. Physical properties

In this segment the investigated different physical properties of spinel-type compounds $CuRh_2S_4$ and $CuRh_2Se_4$ are presented and analyzed with comparison.

*3.1 Structural properties*

The sulfo- and selenospinels $CuRh_2S_4$ and $CuRh_2Se_4$ belongs to the cubic structure with space group $Fd\bar{3}m$ (No.227) [4, 14]. The cubic structure contains 56 atoms in each unit cell. The Wyckoff positions of both these superconductors are 8a (0.125, 0.125, 0.125) for Cu, 16d (0, 0, 0.5) for Rh and 32e (u, u, u) for S or Se, where u is the internal structural parameter [13]. The conventional and primitive unit cells of these spinels have shown in Fig.1. The optimized equilibrium crystal structure of these compounds is received by minimizing the total energy. The investigated structural parameters are recorded in Table 1 with available experimental values. The present investigated parameters are well agreed with experimental values. It is found that the evaluated present lattice parameters (*a*) deviate by 1.01% and 0.65% from experimental values for $CuRh_2S_4$ and $CuRh_2Se_4$. The investigated values and experimental values appears slight difference due to the temperature dependency of cell parameters and GGA process [23]. The variation of lattice parameter, unit cell volume and bulk modulus for $CuRh_2S_4$ and $CuRh_2Se_4$ is appears due to the replacement of atoms by similar atoms which is presented in Fig.2.

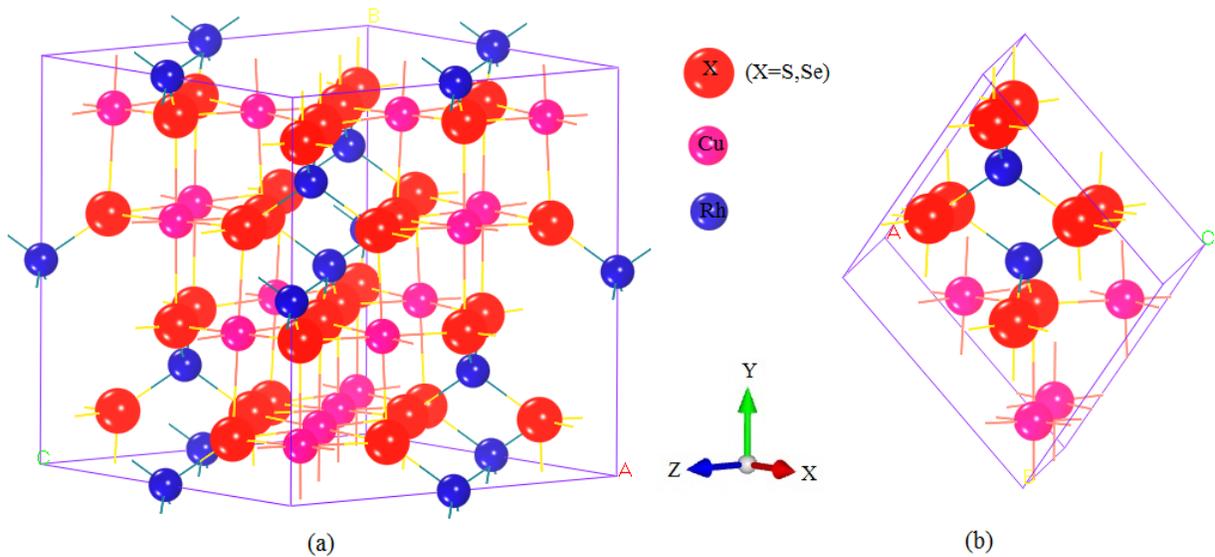

**Fig. 1.** Crystal structure of $CuRh_2X_4$ (where, X=S, Se) (*a*) Conventional unit cell, (*b*) Primitive cell.

**Table 1**
Calculated lattice parameter $a$ (Å), unit cell volume $V$ (in Å$^3$), u (internal structural parameter) and bulk modulus $B$ (in GPa) for CuRh$_2$S$_4$ and CuRh$_2$Se$_4$.

| Compounds | | $a$ | $V$ | $u$ | $B$ | Ref. |
|---|---|---|---|---|---|---|
| CuRh$_2$S$_4$ | Calculated values | 9.887 | 241.69 | 0.366 | 95.66 | This study |
| | Experimental values | 9.787 | 234.36* | 0.384 | - | Expt.[4] |
| | | 9.780 | 233.86* | - | - | Expt.[6] |
| | | 9.790 | 234.57* | - | - | Expt.[14] |
| | | 9.790 | 234.57* | - | - | Expt.[24] |
| | | 9.784 | 234.15* | 0.385 | - | Expt.[25] |
| CuRh$_2$Se$_4$ | Calculated values | 10.336 | 276.12 | 0.366 | 111.50 | This study |
| | Experimental values | 10.269 | 270.72* | 0.384 | - | Expt.[4] |
| | | 10.340 | 276.37* | - | - | Expt.[6] |
| | | 10.270 | 270.80* | - | - | Expt.[14] |
| | | 10.263 | 270.25* | - | - | Expt.[24] |

*Calculated

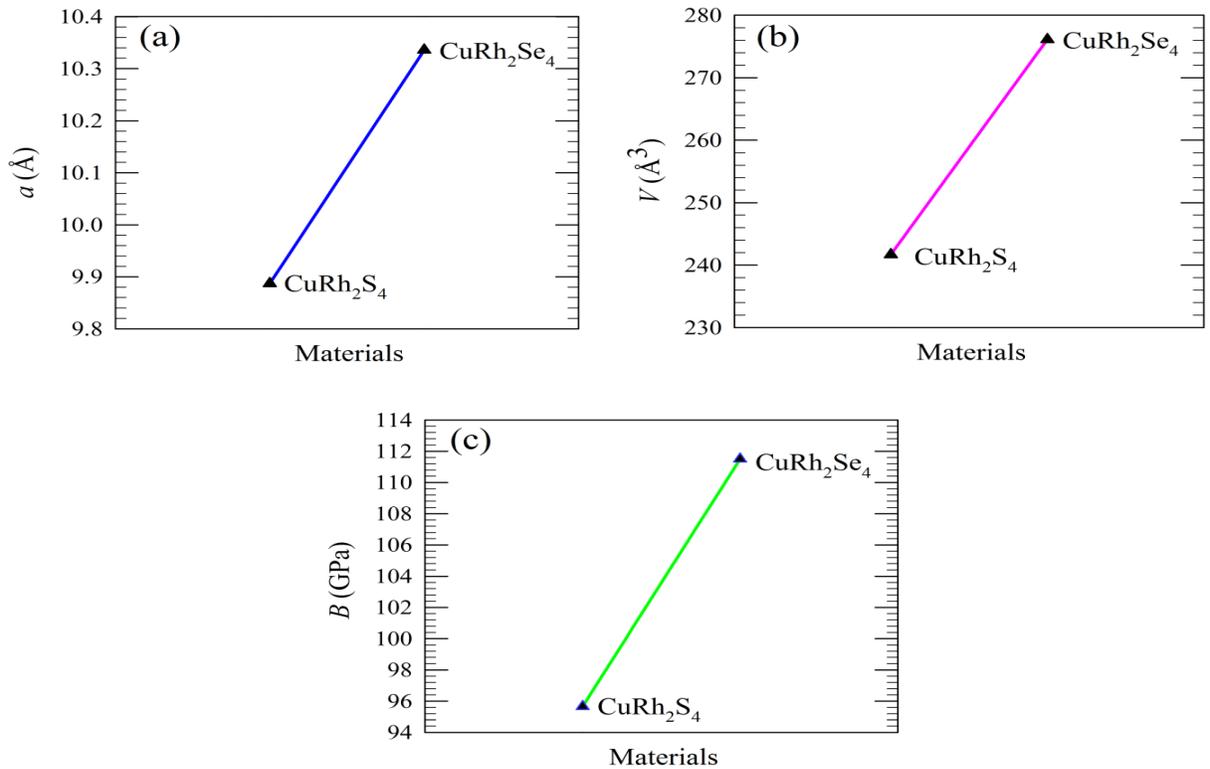

**Fig. 2.** Variation of lattice parameter, unit cell volume and bulk modulus due to the replacement of atoms by similar atoms.

## 3.2 Elastic properties

Elastic constants describe the details mechanical properties and bonding behaviors of solids. Further, those constants use to justify the mechanical stability of solid and connecting the relation with phonon spectrum and Debye temperature of crystals. The elastic properties of any compound nearly related to the long-wavelength phonon spectrum, thus the elastic properties of superconducting material must be investigated [26]. The investigation of the elastic properties further provides frequent knowledge about the dynamical properties of crystalline solids. The elastic constants of crystal have investigated from a linear fit of the evaluated stress-strain function according to the Hook's law [27]. The three independent elastic constants are $C_{11}$, $C_{12}$ and $C_{44}$ for a cubic type crystal structure. For the first time the investigated elastic constant for $CuRh_2S_4$ and $CuRh_2Se_4$ are listed in Table 2. The cubic crystal should be mechanically stable by satisfying the well known Born stability criteria [28]. These criteria are,

$$C_{11} > 0,\ C_{44} > 0,\ C_{11} - C_{12} > 0 \text{ and } C_{11} + 2C_{12} > 0$$

From Table 2, it is evident that the above criteria are satisfied for the present investigated results, which indicating that the $CuRh_2S_4$ and $CuRh_2Se_4$ superconductors are mechanically stable. We are not able to compare present calculated values because no available experimental or other theoretical elastic constants values.

**Table 2**
Calculated elastic constants $C_{ij}$ (GPa), bulk modulus $B$ (GPa), shear modulus $G$ (GPa), Young's modulus $E$ (GPa), $B/G$ ratio, elastic anisotropic factor $A$, Poisson's ratio $v$, Burger's vector $b$ (Å), interlayer distance $d$ (Å), and Peierls stress $\sigma_p$ for $CuRh_2S_4$ and $CuRh_2Se_4$.

| Compounds | $C_{11}$ | $C_{12}$ | $C_{44}$ | B | G | E | B/G | A | v | b | d | $\sigma_p$ |
|---|---|---|---|---|---|---|---|---|---|---|---|---|
| $CuRh_2S_4$ | 129.25 | 94.67 | 30.23 | 106.20 | 25.05 | 69.67 | 4.23 | 1.74 | 0.39 | 9.887 | 4.94 | 0.47 |
| $CuRh_2Se_4$ | 120.78 | 81.51 | 19.47 | 94.60 | 19.53 | 54.81 | 4.84 | 0.99 | 0.40 | 10.336 | 5.16 | 0.35 |

The mechanical properties such as the bulk modulus $B$, shear modulus $G$, Young's modulus $E$ and Poisson's ratio $v$ of $CuRh_2S_4$ and $CuRh_2Se_4$ have calculated by using the Voigt-Reuss-Hill (VRH) equating method [29]. The Voigt and Reuss bounds of $B$ and $G$ have calculated by using the equations for any cubic crystal, as follows [30]:

$$B_v = B_R = \frac{C_{11} + 2C_{12}}{3} \tag{1}$$

$$G_v = \frac{(C_{11} - C_{12} + 3C_{44})}{5} \tag{2}$$

$$G_R = \frac{5C_{44}(C_{11} - C_{12})}{[4C_{44} + 3(C_{11} - C_{12})]} \tag{3}$$

The Hill took an arithmetic mean values of B and G by using the above two paths as follows,

$$B = \frac{1}{2}(B_R + B_v) \tag{4}$$

$$G = \frac{1}{2}(G_v + G_R) \tag{5}$$

Young's modulus ($E$) and Poisson's ratio ($v$) have calculated by using the Hill's bulk modulus ($B$) and shear modulus ($G$) as follows,

$$E = \frac{9GB}{3B + G} \tag{6}$$

$$v = \frac{3B - 2G}{2(3B + G)} \tag{7}$$

The Zener anisotropy factor $A$ have calculated by applying the degree of anisotropy in solid [31] and obtained by the following relation,

$$A = \frac{2C_{44}}{(C_{11} - C_{12})} \tag{8}$$

The evaluated values of the bulk modulus ($B$), shear modulus ($G$), Young's modulus ($E$) and Poisson ratio ($v$), $B/G$ ratio and elastic anisotropic factor $A$ are tabulated in Table 2. From Table 2, it can be seen that the bulk modulus for $CuRh_2Se_4$ relatively low (<100 GPa) and for $CuRh_2S_4$ relatively (>100 GPa) high indicating the soft and hard material by comparing to each other.

The value of bulk modulus is larger than the value of shear modulus ($B>G$) indicating that the shear modulus is the remarkable parameter associating with the stability of $CuRh_2S_4$ and $CuRh_2Se_4$ [32]. Further the larger value of bulk modulus than shear modulus denoting the capacity is stronger of resist deformation. Furthermore shear and Young's modulus indicates the measure of resist reversible deformation by shear stress and stiffness of solid. The calculated values of Pugh's ratio ($B/G$) for $CuRh_2S_4$ and $CuRh_2Se_4$ are found to be 4.23 and 4.84, respectively. The critical value of Pugh's ratio for any material is 1.75, separates the ductile (> 1.75) and brittle (< 1.75) nature of crystals [29]. The present calculated Pugh's ratio for $CuRh_2S_4$ and $CuRh_2Se_4$ indicates the ductile nature. The graphical representations of variation of the Pugh's ratio of spinels have shown in Fig. 3. The critical value of ductility-brittle manner of

Poisson's ratio is 0.33 and above this value the material behaves brittle nature and lower value indicates the ductile nature [33]. The present investigated value indicates the brittle nature. Further, the value of Poisson's ratio gives vital information about the nature of bonding force in solids [34]. The value of $v$ between 0.25 and 0.5 express that the material is central force solid [35]. From Table 2, we see that the value of $v$ is 0.39 and 0.40 for $CuRh_2S_4$ and $CuRh_2Se_4$ denotes the force exists in the material is central. The failure mode of material is known as Cauchy pressure ($C_{12}$-$C_{44}$), when the value is positive and negative then it behaves ductile and brittle nature, respectively [36]. The present value of Cauchy pressure is positive and behaves ductile nature for $CuRh_2S_4$ and $CuRh_2Se_4$. The Pugh's ratio and Cauchy pressure both are similar behavior, indicating the credibility of present DFT base work. The degree of elastic anisotropy measure by the Zener anisotropy factor in solid [31]. The unit value ($A=1$) of anisotropy factor indicating the completely isotropic material and the greater or less value from unity ($A<1$ or $A>1$) denotes the degree of elastic anisotropy of material. The present values are 1.74 for $CuRh_2S_4$, indicating the degree of elastic anisotropy and 0.99 for $CuRh_2Se_4$, indicating the approximately isotropic material.

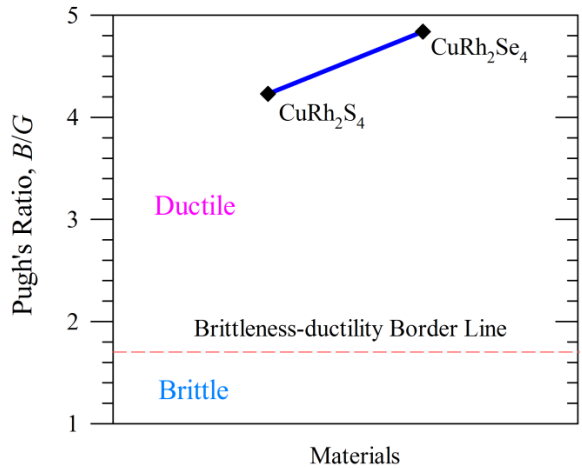

**Fig. 3.** Variation of the Pugh's ratio of spinels $CuRh_2S_4$ and $CuRh_2Se_4$. The horizontal dotted line considers the Border line between Ductile and Brittle.

The Peierls stress of dislocation is the most vital quantity that indicating the strength of a crystal [37] and denotes the force requisite for a dislocation to be in motion within an atomic plane in the unit cell of a crystal. The Peierls stress ($\sigma_p$) can be calculated by the following formula [38].

$$\sigma_P = \frac{G}{1-v}\exp\left(-\frac{2\pi d}{b(1-v)}\right) \qquad (9)$$

Where, $b$ is the Burgers vector and d is the interlayer distance between two glide planes. The calculated values of Burgers vector and interlayer distance with Peierls stress are also recorded in

Table 2. Mirza HK Rubel et al [37] have proved the sequence $(\sigma_p)$ (MAX phases) < $(\sigma_p)$ (new double perovskite) < $(\sigma_p)$ (analogue double perovskite) << $(\sigma_p)$ (binary carbides) for Peierls stress value and several Max phase have the values within the ranges 0.7– 0.98 GPa [39]. They also proved that the dislocations for MAX phase can move easily than perovskite and not possible move for binary carbides. The present calculated values of Peierls stress are 0.47 and 0.35 for spinel compounds $CuRh_2S_4$ and $CuRh_2Se_4$. From the above discussion, it is manifest that the value of Peierls stress of present calculation is less than MAX phase. So, it concludes that the dislocations move very easily for the spinels type compounds.

*3.3 Electronic properties*

In this present study, we have calculated the band structure along the high symmetry directions in the Brillouin zones as well as total and partial density of states. The band structure and DOS closely related with the charge density difference and Fermi surface. For this reason we have also calculated Fermi surface and charge density difference for $CuRh_2S_4$ and $CuRh_2Se_4$. The Fermi surface and charge density difference farther related with the bonding nature and different bonding properties. The calculated electronic band structure is shown in Fig. 4. The horizontal dotted line between the valence band and conduction band considers the Fermi level. From the Figure of band structure it is manifest that the valence bands and conduction bands overlap each other at the Fermi level. Actually, few bands (colored lines) have crossed the Fermi level that found no band gap between valence bands and conduction bands, which also indicates the metallic conductivity of $CuRh_2S_4$ and $CuRh_2Se_4$.

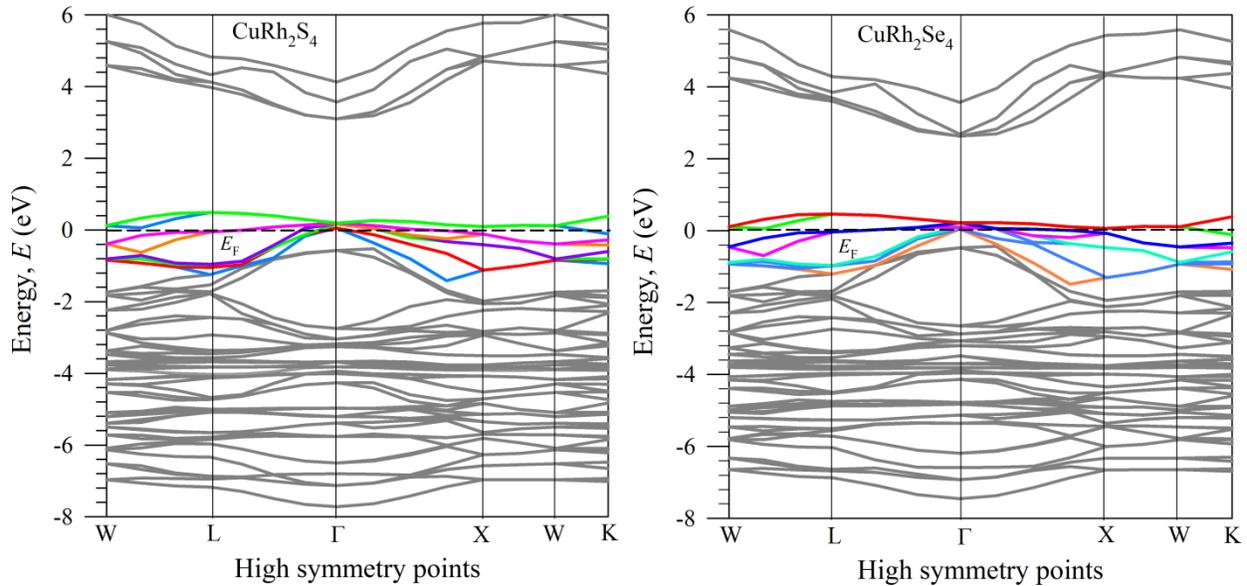

**Fig. 4.** Electronic band structure of (a) $CuRh_2S_4$ and (b) $CuRh_2Se_4$. The horizontal dotted line considers the Fermi level.

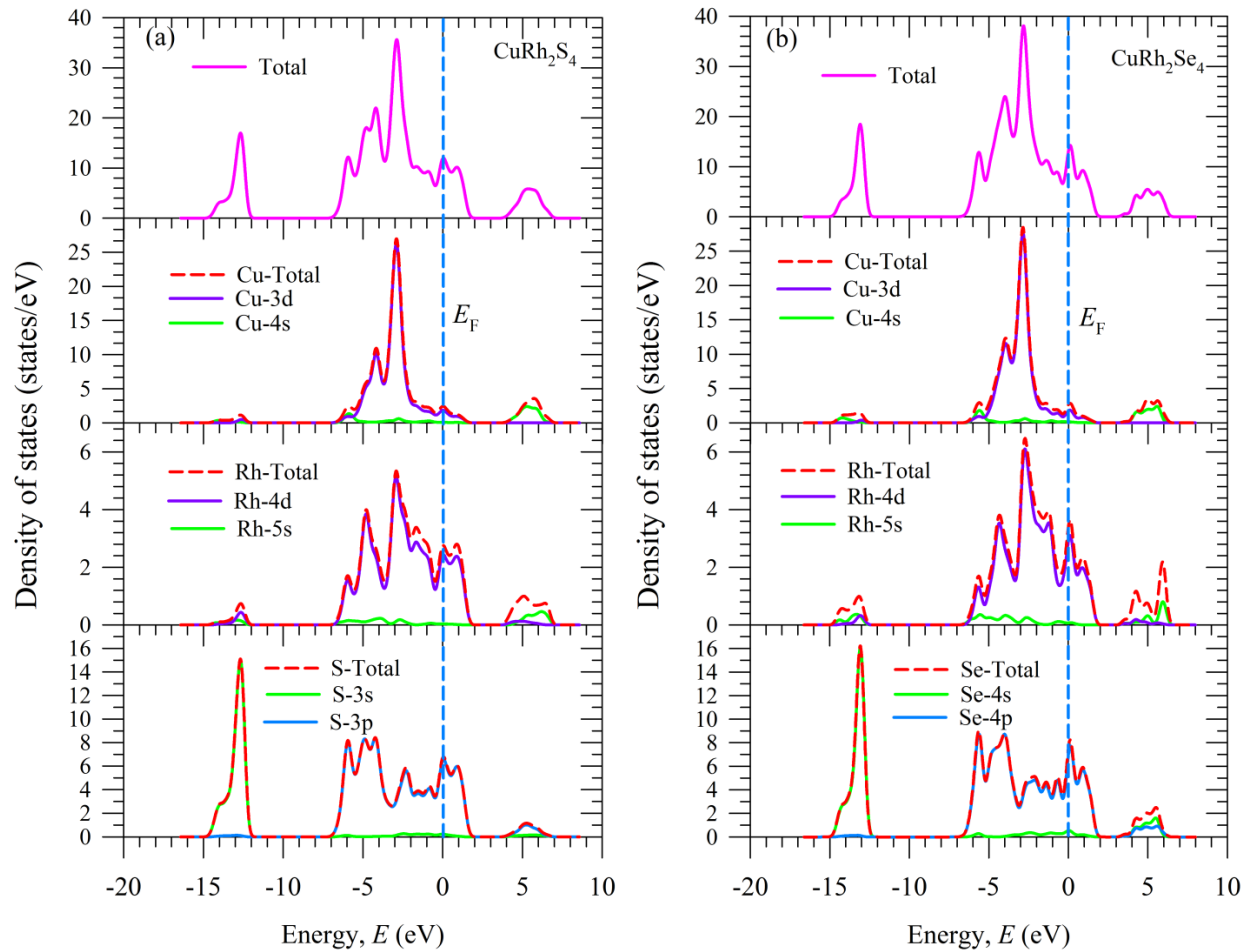

**Fig. 5.** Total and partial density of states of (a) $CuRh_2S_4$ and (b) $CuRh_2Se_4$. The vertical dotted line considers the Fermi level.

The calculated total and partial density of states is shown in Fig. 5. The most contributed state at Fermi level is S-3p rather than Rh and Cu states. The higher peak appears in Cu-3d states at 3eV than Rh and S states. Most of the peaks appears below the Fermi level for these two compounds. The band gaps have found between -3eV to Fermi level energy for Cu and Rh states. The Rh-4d and Cu-3d states for $CuRh_2Se_4$ is more contributed than $CuRh_2S_4$. The overall contribution for $CuRh_2Se_4$ is higher than $CuRh_2S_4$. The calculated total density of states are 13.06 states/eV for $CuRh_2Se_4$ and 11.96 states/eV for $CuRh_2S_4$.

In the contour plot of charge density difference, the collection of charges between two atoms can be made covalents bonds and the ionic bonds can be predicted by the balance of negative and posivite charge at the atom positions [40]. The contour plot of valence electron charge density difference is shown in Fig. 6 along with the 100 crystallographic plane. A scale have shown at the right side (color line) of the contour plot in the units of $e/Å^3$. The red and blue color of scale indicates the high and low electron density. In the plot, the spherical charge distribution appears around S and Se atoms that denote the ionic nature of S-S and Se-Se bonds. The S and Se atoms also highly contributed in the density of states map (Fig. 5) for this reason the charge distribution is maximum for S-S and Se-Se bonds. The ionic character describes the metallic nature of S-S and Se-Se bonds [41]. The electronegativity appears at Cu is relatively high in the charge density difference map because the distribution of charge around Cu is highly uniform and near Cu atoms more charge is accumulate. The covalent bonds nature have found between Cu-Rh bonds because the hybridization found between Cu and Rh atoms. We have also investigated the charge density difference map in different crystallographic plane and have found the same results as well as that found in the present investigation. The charge distribution between S-Cu and Se-Cu are very low due to interatomic distance. Fruther, the lower charge distribution appears between S-Rh and Se-Rh bonds in case of interatomic distance. The overall superior discussion of the charge density differnce map reflects that $CuRh_2S_4$ and $CuRh_2Se_4$ can be designated as an exceedingly anisotropic combination between ionic, covalent and metallic nature.

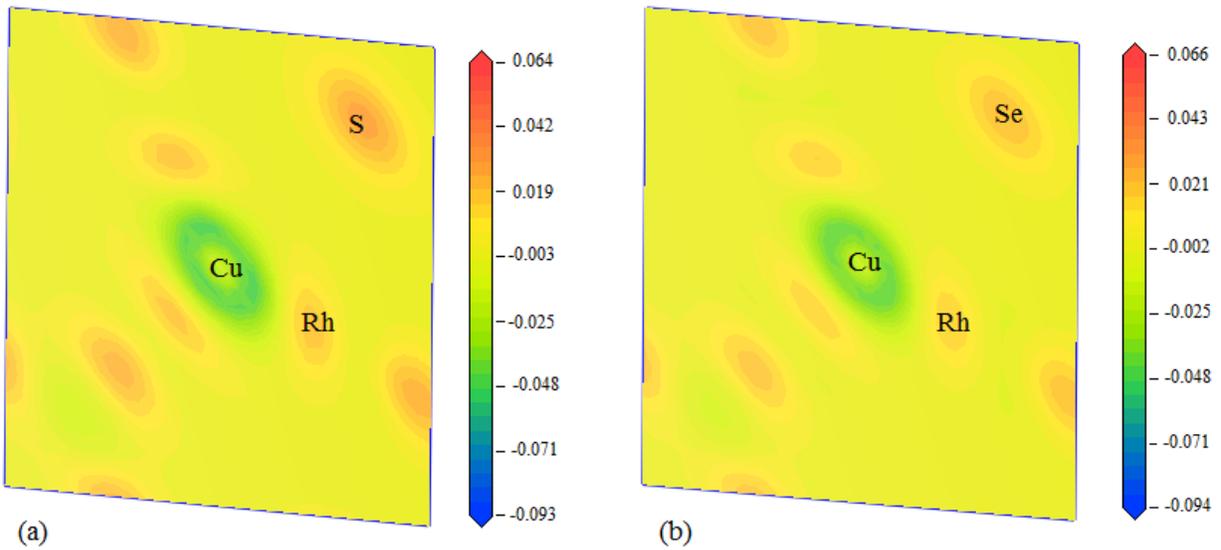

**Fig. 6.** Charge density difference of $CuRh_2S_4$ and $CuRh_2Se_4$ for 100 planes.

The notion of the topology of Fermi surface (FS) can support us to reveals the proper theoretical models concerning the pairing symmetry for any superconducting materials and find the relation between structural and electronic properties [37]. We have plotted the FS topology in Fig. 7 for $CuRh_2S_4$. The FS have calculated with band crossing lines (color lines) in the Fermi level as shown in Fig. 4(a) and have also calculated with the 3D cross section of the Brillouin zone. In the FS topology, we have found eight sheets because of eight band crossing in the Fermi level (as shown in band structure diagram). There are four Fermi-sheets appears at the corner of the topology that indicate hole-like concave sheets at the R-points of the Brillouin zone and another three hole-like sheets found inside the topology at around the Γ point. The spheroid sheet at gamma point in the middle of topology indicates the electron-like sheets [37]. So, overall we conclude that the both electron- and hole-like sheets make the complex multisheet FS of $CuRh_2S_4$. In the present calculation, we are not able to calculate the FS for $CuRh_2Se_4$ due to theoretical process. Hope this will be investigated further in any other theoretical calculation.

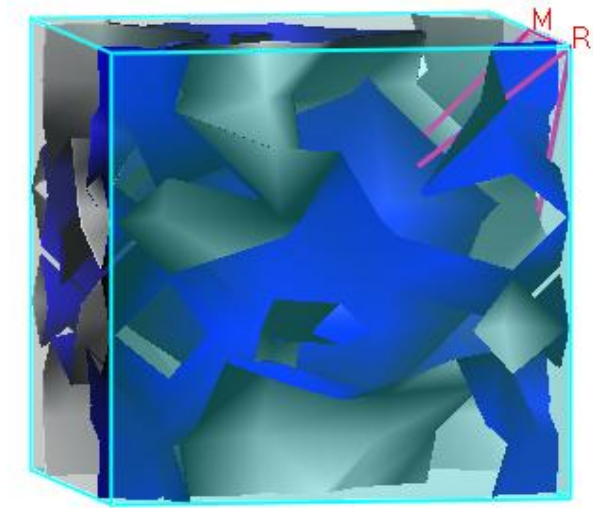

**Fig. 7.** Fermi surface of $CuRh_2S_4$

*3.4 Population analysis and Vickers hardness*

To obtain the different bonding behavior in any crystal system the more investigation is required about the Mulliken atomic populations [42]. The present investigated Mulliken atomic populations of $CuRh_2S_4$ and $CuRh_2Se_4$ superconductors have tabulated in Table 3. In Table 3, we conclude that the S atoms bearing negative charge whereas Cu and Rh bearing positive charge. For the above reason, the charge transfer from Cu and Rh atom to S atom. The charge transformation towards the S atoms from Cu and Rh atoms is equal to 0.05e for $CuRh_2S_4$. For $CuRh_2Se_4$ superconductor, the charge transformation is opposite of $CuRh_2S_4$. The charge is transfer from Se atoms to Cu and Rh atoms which is equal to 0.36e and 0.73e. In the present Mulliken atomic populations S-Rh and S-Cu bonds for $CuRh_2S_4$ behaves high value due to positive population and indicating the increasing level of covalency i.e., highly covalent nature [43]. For $CuRh_2Se_4$ superconductors the Se-Rh and Cu-Se bonds also indicating the covalent nature.

In present investigation, we have also investigated the ionicity of a material ($f_h$) due to understanding the further crucial bonding behavior by using the following equation [44],

$$f_h = 1 - e^{-|P_c - P|/P} \tag{10}$$

Where, $P_C$ represents the bond overlap population in a pure covalent crystal and when the value is one then represents the pure covalent nature whereas P is the bond overlap population. The value of $f_h$ is one (1) and zero (0) indicates the ionic and covalent bond nature of any material. In the present investigation the S-Cu, Se-Rh and Cu-Se bonds represent the relatively ionic nature which contradicts from Mulliken atomic bond nature and only S-Rh bonds relatively covalent nature. This type of contradiction the more investigation is required to overcome.

**Table 3**
Population analysis of CuRh$_2$S$_4$ and CuRh$_2$Se$_4$ superconductors.

| Compounds | Species | s | p | d | Total | Charge | Bond | Population | $f_h$ | Lengths (Å) |
|---|---|---|---|---|---|---|---|---|---|---|
| CuRh$_2$S$_4$ | S | 1.85 | 4.20 | 0.00 | 6.05 | -0.05 | S-Rh | 0.65 | 0.42 | 2.27954 |
| | Cu | 0.67 | 0.64 | 9.60 | 10.90 | 0.10 | S-Cu | 0.36 | 0.83 | 2.39458 |
| | Rh | 0.47 | 0.42 | 8.09 | 8.98 | 0.02 | | | | |
| CuRh$_2$Se$_4$ | Se | 1.53 | 4.11 | 0.00 | 5.64 | 0.36 | Se-Rh | 0.38 | 0.81 | 2.39060 |
| | Cu | 0.80 | 0.92 | 9.64 | 11.36 | -0.36 | Cu-Se | 0.39 | 0.79 | 2.49946 |
| | Rh | 0.78 | 0.78 | 8.16 | 9.73 | -0.73 | | | | |

Hardness evaluates or estimates the resistance to localized plastic deformation initiated by either mechanical indentation or abrasion. Furthermore, evaluating the hardness of a material the amount of force per unit area accomplished the plastic deformation and higher hardness indicates the higher resistance to deformation of a material [45]. The present calculated Vickers hardness have tabulated in Table 4. The Vickers hardness ($H_v$) of a compound acquired from the Mulliken bond population data by using the following relations [46, 47],

$$H_v = \left[ \prod^{\mu} (H_v^{\mu})^{n^{\mu}} \right]^{1/\sum n^{\mu}} \tag{11}$$

$$H_v^{\mu} = 740 \, (P^{\mu} - P^{\mu\prime}) (v_b^{\mu})^{-5/3} \tag{12}$$

$$v_b^{\mu} = (d^{\mu})^3 \Big/ \sum_v [(d^v)^3 N_b^v] \tag{13}$$

$$P^{\mu\prime} = n_{free} / V \tag{14}$$

$$n_{free} = \frac{difference}{3}[1st + 4 \times odd + 2 \times even + Last] \quad (15)$$

Where, $P^\mu$ called the Mulliken population of the $\mu$-type bond, $P^{\mu'}$ denotes the metallic population of the $\mu$-type bond, $v_b^\mu$ is the volume of a bond of type $\mu$, $n_{free}$ is the number of free electrons, $V$ is the cell volume, $d^\mu$ is the bond length of type $\mu$ and $N_b^v$ called the bond number of type $v$ per unit volume.

At present to my knowledge, Diamond is the much hardest material rather than all others materials with range of Vickers hardness 70 to 150 GPa [45]. The present calculated values for CuRh$_2$S$_4$ and CuRh$_2$Se$_4$ superconductors are 2.51 GPa and 1.56 GPa, respectively. With comparing to Diamond, we finally conclude that the present calculated materials are obviously soft material and deformation resistance is low. These materials also have low dimension and suitable to use in wires and ribbon cables.

**Table 4**
Calculated Mulliken bond overlap population of $\mu$-type bond $P^\mu$, n$^\mu$ is the number of bonds, bond length $d^\mu$, metallic population $P^{\mu'}$, bond volume $v_b^\mu$ (Å$^3$) and Vickers hardness of $\mu$-type bond $H_v^\mu$ (GPa) and total hardness $H_v$ (GPa) of CuRh$_2$S$_4$ and CuRh$_2$Se$_4$ superconductors.

| Compounds | bond | $n^\mu$ | $d^\mu$ | $P^\mu$ | $P^{\mu'}$ | $v_b^\mu$ | $H_v^\mu$ | $H_v$ |
|---|---|---|---|---|---|---|---|---|
| CuRh$_2$S$_4$ | S-Rh | 8 | 2.27954 | 0.65 | 0.1612 | 13.9921 | 4.45 | 2.51 |
|  | S-Cu | 8 | 2.39458 | 0.36 | 0.1612 | 16.2203 | 1.42 |  |
| CuRh$_2$Se$_4$ | Se-Rh | 8 | 2.39060 | 0.38 | 0.1422 | 16.1060 | 1.71 | 1.56 |
|  | Cu-Se | 8 | 2.49946 | 0.39 | 0.1422 | 18.4085 | 1.43 |  |

*3.5 Optical properties*

The optical properties of a material describe the interaction nature with light and describe the more about the electronic structure. The optical properties of a matter include the different terms like as dielectric function, refractive index, extinction coefficient, absorption, conductivity, loss function, reflectivity etc. The present calculated optical function for CuRh$_2$S$_4$ and CuRh$_2$Se$_4$ are recorded in Fig. 8 and this calculations stand on up to 25 eV energy range. The value of 0.5 eV Gaussian smearing has elected for all optical properties calculations. The optical properties have calculated by using the frequency dependent dielectric function $\varepsilon$ ($\omega$) = $\varepsilon_1$ ($\omega$) + $i\varepsilon_2$ ($\omega$), Where the imaginary part [$\varepsilon_2$ ($\omega$)] obtained from the momentum matrix elements between the filled and the unfilled electronic eigenstates. The real part of dielectric function [$\varepsilon_1$ ($\omega$)] obtained from the

Kramers-Kronig transform of imaginary part and the real part have defined by the Eqs. 49 to 54 in ref [16]. Further, the imaginary part defined by the following equation [16],

$$\varepsilon_2(\omega) = \frac{2e^2\pi}{\Omega\varepsilon_0} \sum_{k,v,c} |\psi_k^c|u.r|\psi_k^v|^2 \delta(E_k^c - E_k^v - E) \tag{16}$$

Where, $u$ and $\Omega$ are denoted as the polarization of the incident electric field and the unit cell volume, $\omega$ and $e$ are reveals the frequency of light and the charge of electron, $\psi_k^c$ and $\psi_k^v$ are defined as the conduction band wave function and the valence band wave function at $K$ respectively.

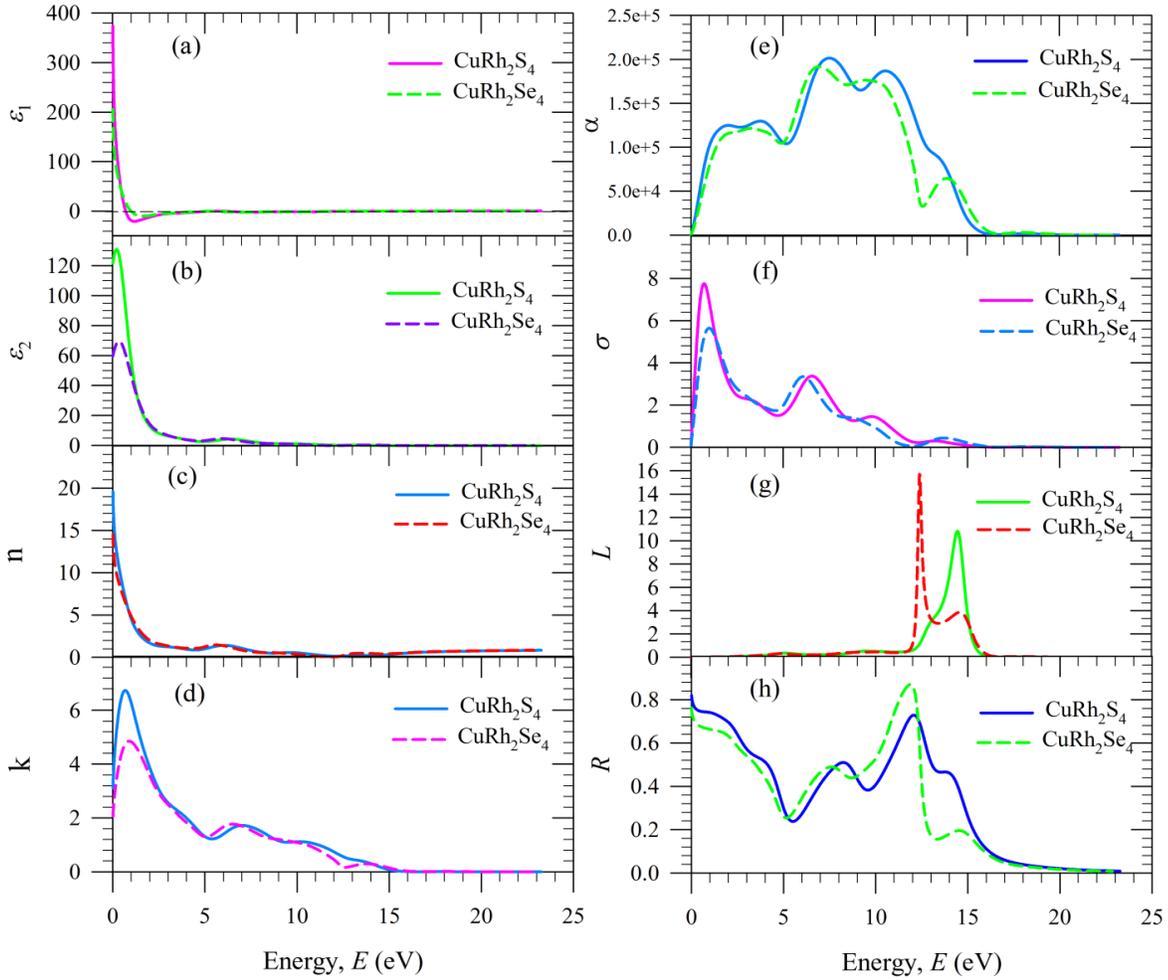

**Fig. 8.** Energy dependent optical functions (*a*) real part of dielectric function, (*b*) imaginary part of dielectric function, (*c*) refractive index, (*d*) extinction coefficient, (*e*) absorption, (*f*) real part of conductivity, (*g*) loss function, and (*h*) reflectivity of $CuRh_2S_4$ and $CuRh_2Se_4$ along [100] polarization directions.

The real and imaginary parts of the dielectric function for both the material are displayed in Fig. 8 (a) and (b). The present calculated dielectric function is high in the initial energy range and decrease continuously when increases the energy range. For real part, the dielectric function stable after 3 eV and for imaginary part the dielectric function stable after 9 eV energy range. The real and imaginary dielectric function is higher for $CuRh_2S_4$ with comparing to $CuRh_2Se_4$. With this comparison, we conclude that $CuRh_2Se_4$ is more transparent than $CuRh_2S_4$. Moreover, $CuRh_2S_4$ is more dielectric, so that this is more suitable to use in a capacitor than $CuRh_2Se_4$.

The real part of refractive index known as the phase velocity which have displaced in Fig. 8(c). The similar contribution found for $CuRh_2S_4$ and $CuRh_2Se_4$ in the whole graph but at zero energy point refractive index value is high for $CuRh_2S_4$. The extinction coefficient (imaginary part) reveals the amount of absorption loss when the electromagnetic wave (as light) passes through the material. The extinction coefficients have also shown in Fig. 8(d). From this graph, it is manifest that the value of extinction coefficient for $CuRh_2S_4$ is higher than $CuRh_2Se_4$. For the reason, $CuRh_2S_4$ is more concentrated than $CuRh_2S_4$. Further, we conclude that $CuRh_2S_4$ strike more electromagnetic wave than $CuRh_2Se_4$.

The absorption coefficient is evaluating the rate of decrease in the intensity of electromagnetic radiation (as light) as it passes via a given substance before it is absorbed. The calculated absorption coefficients are shown in Fig. 8(e). The more effective value for absorption coefficient has found in the visible region and overall contribution for $CuRh_2S_4$ is more than $CuRh_2Se_4$. So, we conclude that $CuRh_2S_4$ is more suitable to use in solar cell rather than $CuRh_2Se_4$.

The present calculated conductivity spectra have displaced in Fig. 8(f) and the effect of conductivity found in the beginning energy region for both materials. The more conductivity is found for $CuRh_2S_4$ than $CuRh_2Se_4$. For the reason, $CuRh_2S_4$ is better to use in electrical conductors rather than $CuRh_2Se_4$.

The loss function represent the loss of energy when fast electron traversing any materials. At which point energy loss appears maximum then this point is called the Bulk plasma frequency and this happen only for the conditions, $\varepsilon_2(\omega)$ is less than one ($\varepsilon_2 < 1$) and $\varepsilon_1(\omega)$ is equal to the zero ($\varepsilon_1 = 0$) [48, 49]. The present calculated loss functions have shown in Fig. 8 (g). The calculated bulk frequencies are 12.00 eV and 14.00 eV for $CuRh_2Se_4$ and $CuRh_2S_4$, respectively. When the incident photon frequency is greater the present materials appears as the transparent materials.

Reflectivity of a material measures the efficiency of a surface to reflect radiation. The present calculated [as shown in Fig. 8(h)] reflectivity represents the good reflectivity in the infrared and visible region. In higher region the less reflectivity appears for both calculated materials. The initial reflectivity for $CuRh_2S_4$ is more effective than $CuRh_2Se_4$ and visible region $CuRh_2Se_4$ is effective than $CuRh_2S_4$. The overall reflectivity has found for both the materials as like as same and both materials have suitable to use as reflector material.

*3.6 Debye temperature*

In a fixed temperature or minimum temperature, the highest frequency mode of vibration is known as Debye temperature. The different physical properties have influence by the Debye temperature such as melting point, specific heat, thermal expansion etc. Debye temperature discuss about the high region of temperature and for the condition $T > \theta_D$ the vibration mode is equal with $K_B T$ energy in every case otherwise vibration mode found at rest. Low temperature region also discuss by the Debye temperature and this mainly comes from acoustic vibration.

The Debye temperature determine only from approximation method because this is not an accurately determined parameter for any material. Various data can be used to approximate the Debye temperature of a material but in present calculation this determine by using the elastic modulus data. The standard method for determination of Debye temperature only depends on the elastic modulus data [50]. The Debye temperature and its related component have calculated by using the following equations [51-54],

$$\theta_D = \frac{h}{k}\left[\frac{3n}{4\pi}\left(\frac{N_A \rho}{M}\right)\right]^{\frac{1}{3}} v_m \qquad (17)$$

$$v_m = \left[\frac{1}{3}\left(\frac{2}{v_t^3} + \frac{1}{v_l^3}\right)\right]^{-\frac{1}{3}} \qquad (18)$$

$$v_l = \left(\frac{B + \frac{4}{3}G}{\rho}\right)^{\frac{1}{2}} \qquad (19)$$

$$v_t = \left(\frac{G}{\rho}\right)^{\frac{1}{2}} \qquad (20)$$

Where, $h$ and $k$ are the Planck constant and Boltzmann constant, $N_A$ is the Avogadro's number, $\rho$ is the density, $M$ is known as the molecular weight and $n$ is the number of atoms in the unit cell of $CuRh_2S_4$ and $CuRh_2Se_4$ superconductors.

The present calculated values of $\rho$, $v_t$, $v_l$, $v_m$ and $\theta_D$ have listed in Table 5. The present evaluated Debye temperatures are 294.23 K and 215.57 K, respectively. The present calculated Debye temperature for $CuRh_2S_4$ contradicts with the experimental value. The contradiction found due to theoretical process and to overcome this contradiction more investigation is required. For $CuRh_2Se_4$ the value of Debye temperature coincides with experimental value.

The calculated phase $CuRh_2S_4$ and $CuRh_2Se_4$ have lower Debye temperature with comparison to a candidate material $Y_4Al_{12}O_9$ for thermal barrier coating [55]. Since lower Debye temperature depends in a lower phonon thermal conductivity, hence $CuRh_2S_4$ and $CuRh_2Se_4$ have a lower thermal conductivity. For the reason, $CuRh_2S_4$ and $CuRh_2Se_4$ should have advantages to use as a thermal barrier coating (TBC) material.

**Table 5**
The calculated density $\rho$ (in gm/cm$^3$), transverse ($v_t$), longitudinal ($v_l$), and average sound velocity $v_m$ (m/s) and Debye temperature $\theta_D$ (K) of $CuRh_2S_4$ and $CuRh_2Se_4$ superconductors.

| Compound | $\rho$ (gm/cm$^3$) | $v_l$ (m/s) | $v_t$ (m/s) | $v_m$ (m/s) | $\theta_D$ (K) | Ref. |
|---|---|---|---|---|---|---|
| $CuRh_2S_4$ | 4.91 | 5332.14 | 2258.72 | 2553.63 | 294.23 | This study |
|  | - | - | - | - | 230 | Expt. [4] |
| $CuRh_2Se_4$ | 6.55 | 4291.65 | 1726.75 | 1955.63 | 215.57 | This study |
|  | - | - | - | - | 211 | Expt. [4] |

*3.7 Thermodynamic properties*

At which temperature the solid change its state to liquid at atmospheric pressure then this phenemenon is know as molting point of solid. The solid and liquid also found in equilibrium at the melting point. Fine et al. proposed a formula for cubic material which has used to calculate the melting temperature, as follows [56],

$$T_m = 553 + 5.91\, C_{11} \qquad (21)$$

Where, the unit of $T_m$ is in K and $C_{11}$ in GPa. The examined melting temperature recorded in Table 6. The melting temperature of $CuRh_2S_4$ is higher with compare to $CuRh_2Se_4$. For the reason, the material $CuRh_2Se_4$ has more convenient effect to melt down than $CuRh_2S_4$.

Thermal conductivity is the essential property of any material that reveals the conduction of heat. The dependency found for thermal conductivity with temperature. The temperature dependent conductivity of any material has increase gradually when temperature decreases to a certain extent [57]. Many different methods have been used for evaluating minimum thermal conductivity ($K_{min}$) but for present investigation we have chosen Clarke method [58]. Clarke method is as follows,

$$K_{min} = K_B \, v_m \left(\frac{M}{n\rho N_A}\right)^{-2/3} \tag{22}$$

Where, $K_B$ and $v_m$ are the Boltzmann constant and the average sound velocity, $M$ is the molecular mass, $n$ is the number of atoms per molecule and $N_A$ is the Avogadro's number.

The calculated temperature dependent minimum thermal conductivity for $CuRh_2S_4$ and $CuRh_2Se_4$ are listed in Table 6. The calculated minimum thermal conductivity of $CuRh_2S_4$ is slightly higher than $CuRh_2Se_4$ but with comparison both are relatively same. For the reason, we conclude that those materials bear low thermal conductivity at ambient conditions.

**Table 6**
The calculated melting temperature, $T_m$ (K), minimum thermal conductivity, $K_{min}$ (in $Wm^{-1}K^{-1}$) and the Dulong-Petit limit (J/mole.K) of $CuRh_2S_4$ and $CuRh_2Se_4$ superconductors.

| Compound | $T_m$ | $K_{min}$ | Dulong-Petit limit |
|---|---|---|---|
| $CuRh_2S_4$ | 1316.86 | 0.49 | |
| | | | 174.55 |
| $CuRh_2Se_4$ | 1266.81 | 0.35 | |

Dulong-Petit limit reveals the anharmonic effect of the specific heat capacity $C_V$ that is suppressed and close to a limit at high temperature [59]. The examined Dulong-Petit limit of a material have found by using the following equation [59],

$$Dulong-Petit \; limit = 3nN_A K_B \tag{23}$$

Where, $N_A$ is the Avogadro's number and $K_B$ is the Boltzmann constant. The calculated value of Dulong-Petit limit has found 174.55 J/mole.K for both the materials.

*3.8 Electron-Phonon coupling constant*

Electron-phonon coupling constant is the most essential superconducting parameter for evaluating superconducting critical temperature. In case of determination of transition temperature electron-phonon coupling constant must be needed to accuracy measurement. For the accuracy measurement of electron-phonon coupling QUANTUM- ESPRESSO program provides good impact [60]. Further, the requirement of double-delta function integration over a dense net of electron and phonon vectors $k$ and $q$ vectors have also needed for accuracy measurement [61].

Moreover, the electron-phonon constant measures accurately by using the experiment value of specific heat coefficient γ. The experiment value of specific heat coefficient γ has also available in literature [4] but we are unable to measure the electron-phonon coupling constant accurately due to theoretical process. Since, we can't measure this accurately, the accuracy of critical temperature ($T_c$) for $CuRh_2S_4$ and $CuRh_2Se_4$ can't be ensured.

In case of above fact the electron-phonon coupling constant have calculated circuitously by using the McMillan formula [62].

$$\lambda_{ep} = \frac{1.04 + \mu^* \ln\left(\frac{\theta_D}{1.45T_c}\right)}{(1 - 0.62\mu^*)\ln\left(\frac{\theta_D}{1.45T_c}\right) - 1.04} \tag{24}$$

Where, $\theta_D$ denotes the Debye temperature and μ* define as coulomb pseudo potential.

Since, coulomb pseudo potential is less sensitive for evaluate the electron-phonon coupling, we take on empirically μ*= 0.13 [62]. The Debye temperatures have used in this section which calculated earlier in present investigation and the experiment transition temperatures are 4.70 K and 3.48 K for $CuRh_2S_4$ and $CuRh_2Se_4$ [4]. After computing this above values, we have found electron-phonon coupling constant $\lambda_{ep}$ = 0.632 ($CuRh_2S_4$) and $\lambda_{ep}$ = 0.634 ($CuRh_2Se_4$). These calculated values well agreed with experiment values in ref. [4]. This good agreement with experimental values indicates the reliability of present calculation.

## 4. Conclusion

In brief, the structural, elastic, electronic, Vickers hardness, vibrational, optical and thermodynamics properties and electron phonon coupling constant of $CuRh_2S_4$ and $CuRh_2Se_4$ have been calculated by using the density functional theory. The optimized structural parameters well agreed with available values. The valence and conduction band overlapped with each other at Fermi level denotes the metallic character of $CuRh_2S_4$ and $CuRh_2Se_4$. The partial density of state shows that S-3p and Se-4p states are more significant at the Fermi level. The mechanical stability found by fulfills the conditions. The Pugh's ratio and Poisson's ratio indicates the ductile and brittle nature, respectively. The bonding properties indicates the ionic, covalent and metallic nature of $CuRh_2S_4$ and $CuRh_2Se_4$. The electron and hole-type sheets appear from the Fermi surface of $CuRh_2S_4$. The value of Vickers-Hardness indicates the soft material with compare to Diamond. The reflectivity spectra indicates that $CuRh_2S_4$ and $CuRh_2Se_4$ have more effect to use a solar reflector. The absorption spectra reveal that $CuRh_2S_4$ behaves good absorber rather than $CuRh_2Se_4$. Debye temperature indicates that these compounds have also more potential to use as a thermal barrier coating (TBC) material. The study of melting temperature of $CuRh_2S_4$ is higher than that of $CuRh_2Se_4$ superconductor and indicates that the materials bear low thermal conductivity at ambient conditions. The melting temperature notifies that the material $CuRh_2Se_4$ has more convenient effect to melt down than $CuRh_2S_4$. The calculated value of Dulong-Petit limit is 174.55 J/mole.K. The electron-phonon coupling constant exhibits that the materials are phonon-mediated medium coupled BCS superconductors. These investigated results could provide great knowledge for other theoretical and experimental spinel-type compounds.